\documentclass[prl,aps,twocolumn, %
showpacs]{revtex4-1}

\usepackage{preamble}

\begin{document}

\date{\today}

\title{Universal Efimov Scaling in the Rabi-Coupled Few-Body Spectrum}

\author{Anthony N. Zulli}
\affiliation{School of Physics and Astronomy, Monash University, Victoria 3800, Australia}
\affiliation{ARC Centre of Excellence in Future Low-Energy Electronics Technologies, Monash University, Victoria 3800, Australia}

\author{Brendan C. Mulkerin}
\affiliation{School of Physics and Astronomy, Monash University, Victoria 3800, Australia}
\affiliation{ARC Centre of Excellence in Future Low-Energy Electronics Technologies, Monash University, Victoria 3800, Australia}

\author{Meera M. Parish}
\affiliation{School of Physics and Astronomy, Monash University, Victoria 3800, Australia}
\affiliation{ARC Centre of Excellence in Future Low-Energy Electronics Technologies, Monash University, Victoria 3800, Australia}

\author{Jesper Levinsen}
\affiliation{School of Physics and Astronomy, Monash University, Victoria 3800, Australia}
\affiliation{ARC Centre of Excellence in Future Low-Energy Electronics Technologies, Monash University, Victoria 3800, Australia}

\begin{abstract}
We investigate the behavior of the Efimov effect---a universal quantum few-body phenomenon---in the presence of an external driving field. Specifically, we consider up to three bosonic atoms, such as  $^{133}$Cs, interacting with a light atom, such as  $^{6}$Li, where the latter has two internal spin states $\{\up$, $\down\}$ that are Rabi coupled.  
Assuming that only the spin-$\up$ light atom interacts with the bosons, we find that the Rabi drive transposes the entire Efimov spectrum such that the Efimov trimers and tetramers are centered around the Rabi-shifted two-body scattering resonance. Crucially, we show that the Rabi drive preserves the trimers’ discrete scaling symmetry, while universally shifting the Efimov three-body parameter, leading to a log-periodic modulation in the spectrum as the Rabi drive is varied. 
Our results suggest that Efimov physics can be conveniently explored using an applied driving field,  opening up the prospect of an externally tunable three-body parameter. 
\end{abstract}
\maketitle

Despite the conceptual simplicity of the quantum few-body problem, it is home to exotic, universal physics. 
A prominent example is the Efimov effect~\cite{efimov1970, efimov1973}, where particles with resonant short-range interactions can host a series of ever larger three-body bound states (trimers).  
Notably, when the particles are on the cusp of forming a bound pair, the two-body scattering length $a$ diverges and the number of trimers tends to infinity. Signatures of Efimov trimers were first observed in a system of identical bosonic atoms~\cite{Kraemer2006}, but have since been reported in mass-imbalanced systems~\cite{barontini2009,pires2014,tung2014} and three-component Fermi gases~\cite{williams2009,Lompe2010}, and even for the first excited state of the $^4$He trimer~\cite{kunitski2015}. 
Importantly, the Efimov effect does not depend on the details of the underlying physical system and is therefore a universal phenomenon that can occur across a broad range of energy scales, from nuclear and particle physics~\cite{Bedaque2000,braaten2006}, to ultracold atoms~\cite{naidon2017}, quantum magnets~\cite{nishida2013}, and even in triple-stranded DNA~\cite{Pal2013}.

A key property of all these manifestations of Efimov physics is the three-body parameter, a short-distance scale which sets the deepest bound Efimov trimer and hence uniquely determines the entire Efimov spectrum due to the trimers’ discrete scaling symmetry~\cite{braaten2006,naidon2017}.  
Specifically, by knowing the scattering length $a_-<0$ at which the deepest trimer unbinds, we also know this parameter for all excited Efimov trimers via the relation $a_-^{(n)} = \lambda^{n} a_-$, where $n=1,2,\dots$ and $\lambda$ is a universal scaling factor. Similarly, at the two-body resonance $a\to \pm \infty$, the trimer energies satisfy $E_0^{(n)} = \lambda^{-2n} E_0$, where $E_0$ is the deepest trimer energy. Remarkably, it has also been shown in the cold-atom context that there is a fixed relationship between the three-body parameter and the underlying short-distance physics, parameterized by the van der Waals range $R_\mathrm{vdW}$, e.g., for the canonical case of three identical bosons, $a_-\simeq-9.5R_\mathrm{vdW}$~\cite{Berninger2011}. This so-called van der Waals universality even holds in the presence of dipolar interactions~\cite{oi2024}. Therefore, an important question is whether this relation can be modified, ultimately yielding a tunable three-body parameter. 

\begin{figure}[b!]
    \centering
    \includegraphics[width=\linewidth]{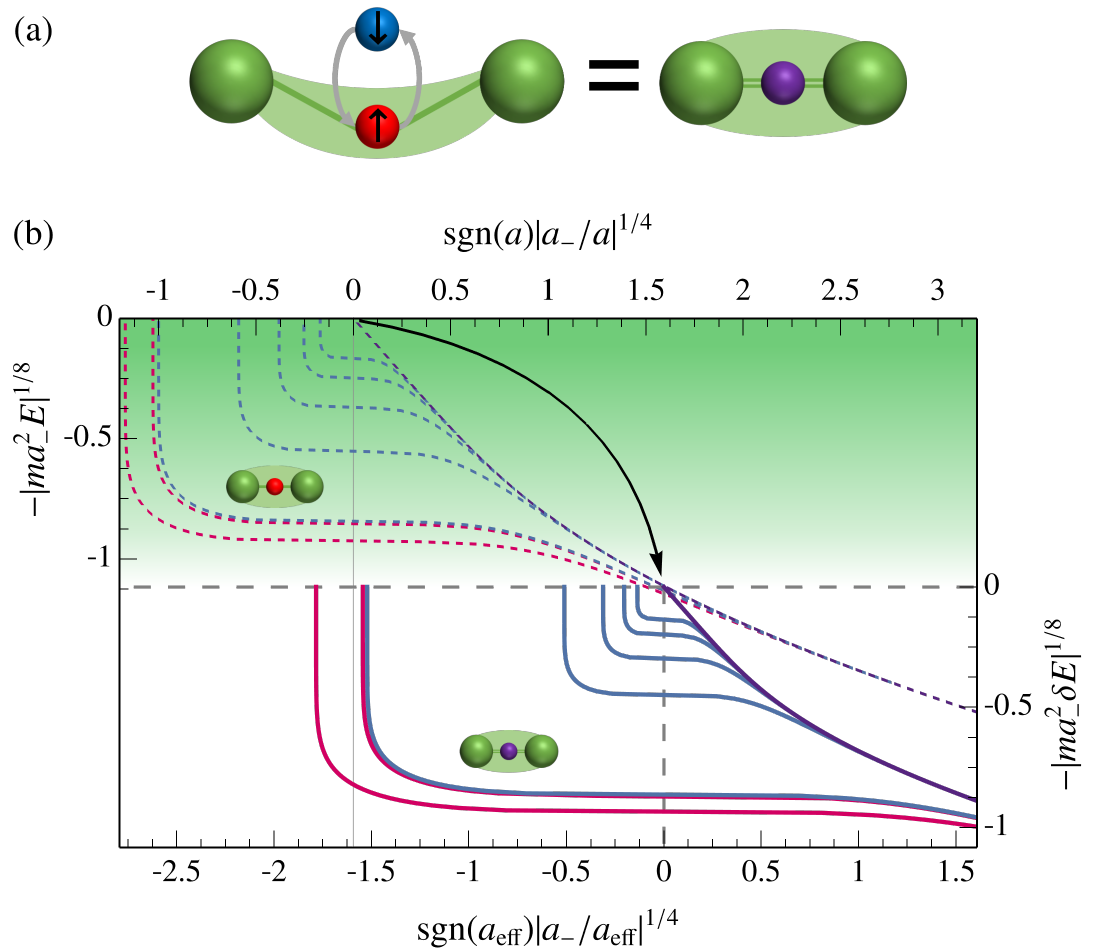}
    \caption{(a) Schematic of the Rabi-coupled three-body problem. The Rabi-dressed atom (purple) consists of a superposition of a spin-$\up$ state (red), which interacts with the heavy bosons (green), and a non-interacting spin-$\down$ state (blue).
    (b) The Efimov spectrum with (solid lines) and without (dashed) the Rabi drive for the mass ratio $133/6$. The axes represent the original parameters (left and top) and the effective parameters in the Rabi-driven case (bottom and right), corresponding to the effective scattering length $a_{\rm eff} \propto (1/a - 1/a_{\rm c})^{-1}$ 
    and the energy $\delta E = E - \es{-}$ measured from the Rabi-shifted continuum. The uncoupled Efimov spectrum is centered around $(1/a=0,E=0)$, while the coupled spectrum is shifted (arrow) to be centered around $(1/a=1/a_\text{c}, E=\es{-})$. 
    The two-body bound state (purple), trimers (blue) and tetramers (pink) are visible in both spectra. The Rabi-coupled Efimov spectrum is calculated for $ma_-^2\Omega_0 = 2$ and $\Delta_{0}/\Omega_0 = -1$.}
    \label{fig:both-spectrum}
\end{figure}

In this Letter, we theoretically show that the Efimov spectrum in an atomic gas can be manipulated with an applied Rabi drive, allowing one to go beyond the usual constraints imposed by atomic physics. 
We consider the scenario of a very light impurity atom interacting with two or more heavy bosonic atoms [Fig.~\ref{fig:both-spectrum}(a)], which is advantageous for accessing the Efimov spectrum due to the smaller scaling factor $\lambda$~\cite{braaten2006, naidon2017}. 
We further assume that the light atom has two internal spin states $\{\uparrow, \downarrow\}$ which are Rabi coupled, e.g., via a radiofrequency driving field, and where only the spin-$\up$ state strongly interacts with the bosons. Similar Rabi-coupled systems have already been successfully implemented experimentally in two-component Bose gases~\cite{shibata2019, guan2020, lavoine2021, sanz2022, cominotti2023,zenesini2024} and, most recently, in highly imbalanced Fermi gases~\cite{vivanco2023}. 

As shown in Fig.~\ref{fig:both-spectrum}(b), we find that the Rabi drive shifts the spectrum of Efimov trimers, and associated four-body bound states (tetramers), such that it lies around the new two-body resonance~\cite{mulkerin2024,bleu2025} for the Rabi-dressed impurity. In particular, we find that the infinite tower of trimer states, along with the usual discrete scaling symmetry, is preserved by the Rabi coupling, unlike the related scenario of spin-orbit coupling~\cite{shi2014,shi2015, guan2018}, where there is instead a modified scaling law in a higher dimensional space~\cite{guan2018}. 
We also show that the whole spectrum can be conveniently explored by fixing the scattering length $a$ and instead varying the detuning of the Rabi drive from the atomic transition. Crucially, we find that the three-body parameter is universally shifted by the Rabi drive, with a log-periodic dependence on the strength of the Rabi coupling.

\paragraph{Model.---}
The Hamiltonian for our Rabi-coupled few-body system consists of three terms: 
\begin{equation}
    \hat{H} = \hat{H}_0 + \hat{H}_\Omega + \hat{H}_\uparrow,
\end{equation}
describing the non-interacting system, the Rabi driving, and the impurity-boson interactions, respectively.
Here, 
\begin{align}
\hat{H}_0 = \sum_{\bk}\epsilon_{\bk}^{\pd b} \oppd{b}_{\bk} \opp{b}_{\bk} + \sum_{\bk\sigma}\epsilon_{\bk}^{\pd c} \oppd{c}_{\bk \sigma} \opp{c}_{\bk \sigma},
\end{align}
where $\oppd{b}_\bk$ ($\oppd{c}_\bk$) creates a heavy (light) particle with momentum $\bk$, $\epsilon_{\bk}^{b} = |\bk|^2/2M \equiv k^2/2M$ is the dispersion for the identical bosons with mass $M$, and $\epsilon_{\bk}^{c} = k^2/2m$ is the corresponding dispersion for the light impurity with mass $m$ and spin $\sigma \in \{\uparrow, \downarrow\}$.
We work in units such that $\hbar$ and the system volume are both set to 1.
We focus on the experimentally accessible case of a $^{133}$Cs-$^6$Li mixture~\cite{tung2014, pires2014, lippi2024}, where the mass ratio $M/m = 133/6$, resulting in an Efimov scaling factor of $\lambda = 4.87$~\cite{braaten2006, naidon2017} when the boson-boson interactions are negligible. For simplicity, we also assume that the spin-$\down$ state is completely non-interacting, but this condition can straightforwardly be relaxed~\cite{adlong2020, hu2023, bleu2025, mulkerin2025}.

We model the coupling of the two impurity spin states within the rotating wave approximation:
\begin{equation}
    \hat{H}_{\Omega} = \frac{\Omega_{0}}{2}\sum_{\bk}(\oppd{c}_{\bk \uparrow} \opp{c}_{\bk \downarrow} + \oppd{c}_{\bk \downarrow} \opp{c}_{\bk \uparrow}) + \Delta_{0} \sum_{\bk}\oppd{c}_{\bk \downarrow} \opp{c}_{\bk \downarrow}\,.
\end{equation}
Here, $\Omega_0$ is the Rabi coupling and $\Delta_{0}$ is the detuning from the bare $\uparrow$-$\downarrow$ transition. Solving the single-particle problem yields two Rabi-split quasiparticle branches with dispersions $E_\bk^\pm = \epsilon_{\bk}^{c} + \es{\pm}$, where $\es{\pm} = \big(\Delta_{0} \pm \sqrt{\Omega_0^2 + \Delta_{0}^2}\big)/2$, for the upper and lower branches, respectively. Their respective (real) spin-$\up$ amplitudes $v$ and $u$ satisfy $u^{2} = \big(1 + \Delta_{0}/\sqrt{ \Omega_{0}^{2} + \Delta_{0}^{2}}\big)/2$
and $u^2 + v^2 = 1$. 

The term describing interactions between the bosons and the spin-$\uparrow$ impurity takes the form
\begin{equation}
   \hat{H}_{\uparrow} = \sum_{\bk} (\epsilon_{\bk}^{\pd d}+\nu_0^\pd) \oppd{d}_{\bk} \opp{d}_{\bk} +g\sum_{\bk\bq} \left( \oppd{d}_{\bq} \opp{c}_{\bq-\bk \uparrow} \opp{b}_{\bk} + \oppd{b}_{\bk} \oppd{c}_{\bq-\bk \uparrow}  \opp{d}_{\bq}\right),
\end{equation}
corresponding to a two-channel model~\cite{timmermans1999} in which two atoms can form a closed-channel dimer $\oppd{d}_\bk$ with kinetic energy $\epsilon_{\bk}^{d} = k^2/2(m + M)$ and detuning $\nu_0$ from the open channel. This model has the exact low-energy scattering amplitude $f(k) = -1/ (a^{-1}+ R^{*}k^{2} + ik)$,
where the scattering length $a$ and range parameter $R^{*}$ replace the bare model parameters via $m_r/2\pi a = -\nu_0/g^2 + \sum_{\bk}1/(\epsilon^b_\bk + \epsilon^c_\bk)$ and $R^{*} = \pi/m_r^2g^2$, where $m_r = mM/(m + M)$ is the reduced mass. The introduction of the closed channel is useful, as it provides a UV cutoff that regulates the Efimov physics, thus introducing an effective three-body parameter. Indeed, for the specific case of a $^{133}$Cs-$^6$Li mixture, the deepest bound Efimov trimer unbinds at the scattering length $a_-=-1.04R^*$ and has energy $E_0 = -0.242/mR^{*2}$ at unitarity.
However, as explicitly shown in the Supplemental Material (SM)~\cite{SM}, our results in the following do not sensitively depend on the choice of UV cutoff~\footnote{Our model aims to capture the essential physics of Rabi-coupled mass-imbalanced mixtures. For direct comparison with Cs-Li experiments, one may need to employ more complicated models in order to capture overlapping Fano-Feshbach resonances~\cite{repp2013,tung2013}, such as models involving three interaction channels~\cite{johansen2017,yudkin2021,li2022}.} or on whether intraspecies scattering is included~\cite{ulmanis2016a}.

\paragraph{Rabi-coupled few-body states.---} We first investigate the two-body bound state which, for $R^*=0$, was already considered in Ref.~\cite{mulkerin2024}. We can obtain the bound-state energy from the \sch equation for the two-body wave function~\cite{SM} or, equivalently, from 
the two-body $T$ matrix~\cite{mulkerin2024}, which describes the scattering of a spin-$\uparrow$ light atom with a heavy atom at energy $E$ and center-of-mass momentum $\bk$, 
\begin{equation} \label{eq:driven-two-body}
    \begin{aligned}
         &\mathcal{T}_{\uparrow}^{-1}(E, \bk) = \frac{m_{r}^2R^{*}}{\pi}(E - \epsilon_{\bk}^{d}) + \frac{m_{r}}{2\pi a} \\ &-\frac{m_{r}^{3/2}}{\sqrt{ 2 }\pi} \left(u^{2}\sqrt{\es{-} + \epsilon_{\bk}^{d} - E} + v^{2}\sqrt{\es{+} + \epsilon_{\bk}^{d} - E}\right)\,.
    \end{aligned}
\end{equation}
This explicitly accounts for the fact that the Rabi-coupled impurity  
explores both spin states during the interaction~\cite{SM}. In the center-of-mass frame, where $\k=0$, the two-body bound state corresponds to a pole below the scattering continuum, i.e., for $E<\es{-}$. Notably, the bound state only exists above a critical scattering length $1/a > 1/a_\text{c}$, as illustrated in Fig.~\ref{fig:both-spectrum}(b). Explicitly, 
\begin{equation}
    \frac{1}{\ac} = v^{2}\sqrt{ 2m_{r}(\es{+} -\es{-})} -2R^{*}\es{-}m_{r}\,.
    \label{eq:critical-scattering}
\end{equation}
Furthermore, following Ref.~\cite{mulkerin2024}, we transform to the upper/lower dressed basis to obtain the $T$ matrix for the lower Rabi-split state: $\mathcal{T}_{\smallminus} = u^2\mathcal{T}_{\uparrow}$.
This has an effective scattering length, given by~\cite{SM}
\begin{equation}
    \begin{aligned} 
        \aef &= u^{2}\left( \frac{1}{a} - \frac{1}{\ac} \right)^{-1} \,,
    \end{aligned}
    \label{eq:scat-param}
\end{equation}
where $\aef$ diverges at $a=\ac$, as expected. Note that this differs from the resonances induced by sinusoidally driving the scattering length~\cite{smith2015}, since there is no imaginary part to $\aef$. There also exists an effective scattering length for the upper Rabi-split state; however this is always complex and does not feature resonances~\cite{mulkerin2024, bleu2025}.

We now go beyond the two-body system, revealing how the Efimov three-body problem is modified by the Rabi drive. The corresponding bound states are found by solving~\cite{SM}
\begin{equation}\mathcal{T}_{\uparrow}^{-1}(E - \epsilon_{\bk}^{b}, \bk)\gamma_{\bk} = \sum_{\bq} \gamma_{\bq} \left( \frac{u^{2}}{E - E^{-}_{\bk \bq}} + \frac{v^{2}}{E - E^{+}_{\bk \bq}} \right)\,,
\label{eq:three-body-T}
\end{equation}
where $E^{\pm}_{\bk_{1}\bk_{2}} = \epsilon_{\bk_{1}}^{b} + \epsilon_{\bk_{2}}^{b} + E_{-\bk_{1}-\bk_{2}}^{\pm}$ and $\gamma_\bk$ is a three-body vertex function. Similar to the standard three-body problem, we find that the solutions of  Eq.~\eqref{eq:three-body-T} yield a spectrum with infinitely many three-body bound states. 
As shown in Fig.~\ref{fig:both-spectrum}(b), these are centered around the new resonance and become unbound at the shifted continuum. 
In order to directly compare with the usual Efimov scenario, we plot the Rabi-coupled spectrum in terms of the effective scattering length, Eq.~\eqref{eq:scat-param}, and measure energy with respect to the shifted continuum, $\delta E = E - \es{-}$. We see that the spectrum resembles the undriven case, with some modifications for the tightest bound states due to the specific drive parameters chosen, which positions the shifted continuum $\es{-}$ between the lowest- and second-lowest-energy trimers, %
effectively shifting these states differently. %
In particular, we find that the excited states satisfy the usual scaling relations for the critical scattering lengths at which the Efimov trimers unbind, $a_{\mathrm{eff},-}^{(n)} = \lambda a_{\mathrm{eff},-}^{(n-1)}$, and  for their energies at unitarity, $\delta E^{(n)} = \lambda^{-2} \delta E^{(n-1)}$. 

\begin{figure}[t!]
    \centering
    \includegraphics[width=\linewidth]{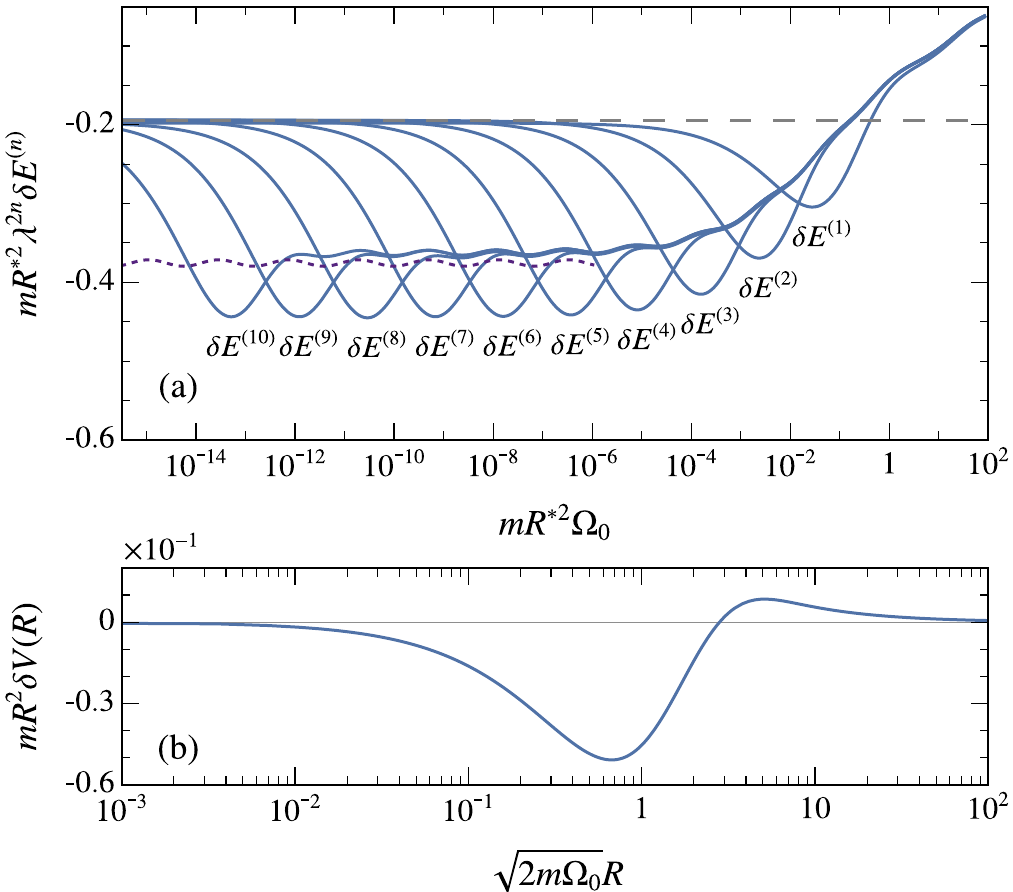}
    \caption{(a) Tuning the three-body parameter via a Rabi drive.  
    We show the energies (solid lines) of the first ten excited Efimov trimers at the induced resonance ($1/a = 1/\ac$) and at $\Delta_0 = 0$ as a function of Rabi-coupling strength, ignoring the less universal ground state. Here, we have scaled the energies $\delta E^{(n)}$ by $\lambda^{2n}$ so that the lines collapse in the asymptotic large-$n$ limit, where the gray dashed line corresponds to the behavior in the absence of Rabi coupling: $E_0^{(n)}\lambda^{2n}=-0.194/mR^{*2}$. The purple dashed line is the Rabi-induced universal behavior obtained from the Born-Oppenheimer (BO) approximation~\cite{SM}. (b) The difference $\delta V(R)$ between the coupled and uncoupled BO potentials at $1/a_{\mathrm{eff}}=0$ and $1/a=0$, respectively, as a function of the distance $R$ between the two heavy atoms. Here, we take $\rs=0$ for both potentials and $\Delta_0 = 0$ for the coupled potential.}
    \label{fig:E0-drive}
\end{figure}

We can also determine the existence of four-body bound states by solving the equation~\cite{SM}
\begin{equation}
    \begin{aligned}&\mathcal{T}_{\uparrow}^{-1}(E - \epsilon_{\bk_{1}}^{b} - \epsilon_{\bk_{2}}^{b}, \bk_{1} + \bk_{2})\gamma_{\bk_{1}\bk_2} =  \\ 
         &\sum_{\bq} (\gamma_{\bk_{1}\bq} + \gamma_{\bq\bk_{2}})\left(  \frac{u^{2}}{E - E^{-}_{\bk_{1} \bk_{2} \bq}} + \frac{v^{2}}{E - E^{+}_{\bk_{1} \bk_{2} \bq}}\right), \end{aligned}
\label{eq:four-body-T}
\end{equation}
where $\gamma_{\bk_{1}\bk_{2}}$ is the four-body vertex and $E^{\pm}_{\bk_{1}\bk_{2}\bk_{3}} = \epsilon_{\bk_{1}}^{b} + \epsilon_{\bk_{2}}^{b} + \epsilon_{\bk_{3}}^{b} + E_{- \bk_{1} - \bk_{2} -\bk_{3}}^{\pm}$. 
As shown in Fig.~\ref{fig:both-spectrum}(b), we find two tetramer states below the ground-state trimer, which mirrors the situation in the absence of a Rabi drive~\cite{ferlaino2009,blume2014}.
This again demonstrates how the Rabi coupling allows us to preserve the main features of the Efimov spectrum, including the associated higher-body bound states.

\paragraph{Dressing the three-body parameter.---}
A particularly intriguing question is whether one can tune the effective three-body parameter of the Efimov spectrum. This parameter is at the heart of the Efimov effect, and is associated with breaking the apparent continuous scale invariance of the resonant three-body problem down to a discrete scaling symmetry~\cite{braaten2006,naidon2017}. In other words, the properties of the weakly bound Efimov spectrum follow from the universal scaling relations, with all details of the UV physics captured in a single parameter. In the cold-atom context, the three-body parameter is typically fixed since it is universally related to the underlying van der Waals interactions~\cite{Berninger2011} due to an effective three-body quantum reflection~\cite{Wang2012Origin,wang2012}.

To investigate the effect of a Rabi drive, we show in Fig.~\ref{fig:E0-drive}(a) the energies of the excited trimers at the induced resonance as a function of the Rabi coupling and at $\Delta_0=0$. In order to directly compare the trimer energies, we have scaled the energies $\delta E^{(n)}$ by $\lambda^{2n}$. 
We see that a given trimer state is only marginally affected when $\Omega_0\ll |E_0^{(n)}|$, in the sense that its binding energy is essentially the same as in the uncoupled system at unitarity: $\delta E^{(n)}\simeq E_0^{(n)}$. 
However, when $\Omega_0\sim |E_0^{(n)}|$ the state becomes deeper bound, only to approach a near-constant log-periodic function when $\Omega_0\gtrsim |E_0^{(n)}|$, where the bound state is roughly twice as tightly bound as in the absence of a Rabi drive. Furthermore, we clearly observe that all the highly excited states where $|E_0^{(n)}|\lesssim \Omega_0$ collapse onto the same log-periodic function of $\Omega_0$, thus indicating that they have the same Rabi-dressed three-body parameter. Crucially, we find that the new three-body parameter is universally shifted from the original three-body parameter by the Rabi drive in a model-independent manner, as explicitly shown in the SM~\cite{SM}.
The possibility of modifying the three-body parameter has also been theoretically explored for the case of a sinusoidally driven scattering length, but no universal behavior was found~\cite{sykes2017}.

We can understand the emergence of a modified three-body parameter %
through the Born-Oppenheimer (BO) approximation. Here, we assume that the two heavy atoms move slowly in comparison to our light particle, allowing us to obtain an effective interaction potential between the two heavy particles~\cite{fonseca1979}. The potential for the uncoupled Efimov system at unitarity is well known, $V_0(R) = -\W(1)^2/2mR^2$, where $R$ is the distance between the two heavy atoms (we assume $R$ is far outside the UV region) and $W$ is the Lambert $W$ function~\cite{fonseca1979}, with $\W(1) \approx 0.567$. However, with the addition of a Rabi drive, the coupled effective potential is more complex~\cite{SM}. 
Of particular interest is the potential at the Rabi-induced resonance: Its deviation $\delta V(R)$ from the universal form $V_0(R)$ above is shown in Fig.~\ref{fig:E0-drive}(b) for the case where $\Delta_0=0$. Crucially, we find that deviations fall off faster than $1/R^2$ at both short and long range. Therefore, while the coupling alters the potential for finite $R \sim 1/\sqrt{m\Omega_0}$, the Rabi-coupled effective potential retains the universal discrete scaling symmetry for both deeply bound and highly excited states~\cite{braaten2006}. 

\begin{figure}[t!]
    \centering
    \includegraphics[width=\linewidth]{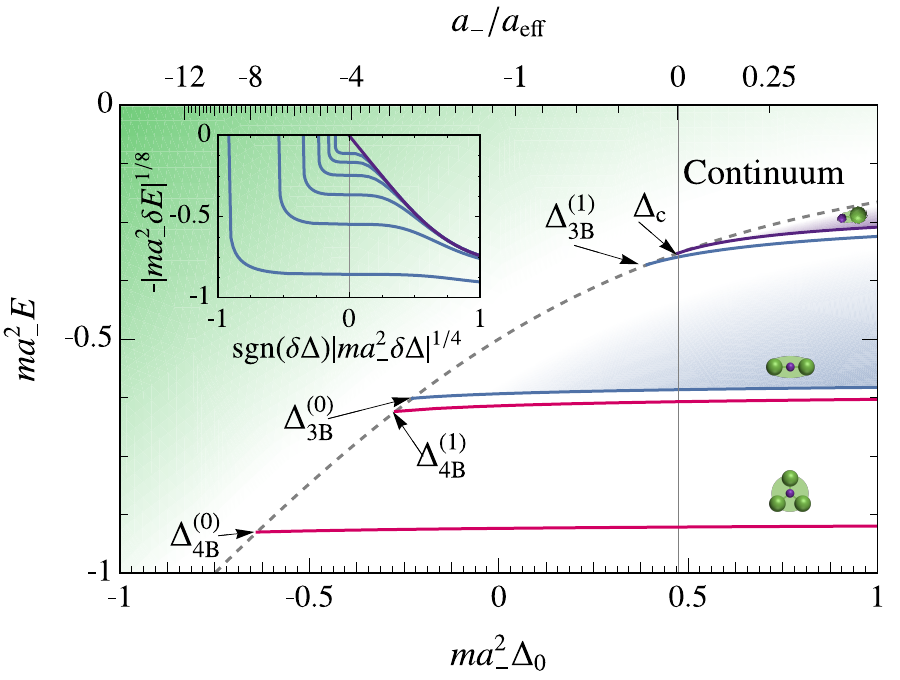}
    \caption{The few-body Efimov spectrum in terms of the detuning and effective scattering length for $ma_-^2\Omega_0 = 1$ and $a=|a_-|$. The $\aef$ (top) axis represents the value of the effective scattering length at the given detuning value [see Eq.~\eqref{eq:scat-param}]. Here, we have shaded the specific regions of interest, the scattering continuum (green), the continuum above the dimer (purple) and the continuum above the ground state trimer (blue). The two-body bound state (purple), trimer states (blue) and tetramer states (pink) all enter the scattering continuum at the lower Rabi-coupled state, with energy $\es{-}$ (dashed line) Inset: The Efimov spectrum centered around the critical detuning $\Dtwo$ as a function of $\delta\Delta = \Delta_{0} - \Dtwo$.}
    \label{fig:delta-spectrum}
\end{figure}

Importantly, while the deeply bound states are essentially unmodified, the higher excited states can, in principle, have a modified three-body parameter. We can determine this by mapping the problem onto an effective one-dimensional scattering problem~\cite{sykes2017} for the potential $\delta V(R) R^2$ shown in Fig.~\ref{fig:E0-drive}(b). 
According to the variable phase approach, the phase $\delta(R)$ accumulated through the potential satisfies~\cite{SM}
\begin{align}\label{eq:variablephase}
s_0\,\frac{d \delta}{d \ln R} = - MR^2\delta V(R) \sin^2[s_0\ln(k_\Omega R) + \delta(R)],
\end{align}
with $s_0\simeq W(1)\sqrt{M/2m_r}$ and $k_\Omega=\sqrt{2m_r\Omega_0}$. For a given Efimov trimer, the ratio of binding energies with and without the Rabi drive is given by $\exp\left(2[\delta(R\to\infty) - \delta(R_0)]/s_0\right)$ where $R_0$ is a short-range scale~\cite{SM}. Applying this analysis directly yields the purple line in Fig.~\ref{fig:E0-drive}(a), which all trimers with $|E_0^{(n)}|\ll \Omega_0$ will follow. This matches our exact numerics very well, with small deviations arising from the use of the BO approximation. We have thus demonstrated that the Rabi-dressed three-body parameter is a log-periodic function of the Rabi coupling, and hence that the three-body parameter is tunable via an external drive. The range of possible values is further increased by considering also the role of the detuning~\cite{SM}.

Equation~\eqref{eq:variablephase} also naturally allows us to consider the modification of inelastic losses due to the Rabi coupling by taking $\delta(R_0)\to\delta(R_0)+i\eta$ where $\eta$ is the inelasticity parameter of the uncoupled system. We find that, like the energy, the Rabi-coupled inelasticity parameter is log-periodic
with increasing $\Omega_0$, and that there are only small oscillations around the bare
parameter~\cite{SM}. Therefore, we expect that the Rabi drive would only minimally affect the
collisional losses.

\paragraph{Detuning as a scattering parameter.---}
Finally, we demonstrate how the entire Efimov spectrum can be explored via the parameters of the Rabi drive and thus, in principle, without requiring tunable Fano-Feshbach resonances. Figure \ref{fig:delta-spectrum} highlights the idea: By fixing the scattering length to a value away from unitarity, in this case $a = |a_-|$, and the Rabi-coupling strength, $ma_-^2\Omega_0 = 1$, we can tune the spectrum by varying the detuning $\Delta_0$. Indeed, we still see the existence of the two tetramers below the trimers, and we find that the trimers have an accumulation point at the critical value of detuning $\Delta_c$ at which the effective scattering length $a_\mathrm{eff}$ diverges. The latter can be understood from the fact that in the vicinity of $\Delta_{0} = \Dtwo$ we have $\aef (\Delta_{0}) \propto 1/(\Delta_{0} - \Dtwo)$. Therefore, the spectrum evaluated in terms of $\delta\Delta = \Delta_{0} - \Delta_c$ and $\delta E$ will mirror the uncoupled spectrum in Fig.~\ref{fig:both-spectrum}, as explicitly shown in the inset of Fig.~\ref{fig:delta-spectrum}.

\textit{Discussion.---} 
Experimentally, one can probe the Efimov spectrum under a continuous Rabi drive by starting in the non-interacting state at $\Delta_0/\Omega_0 \ll -1$ and then adiabatically varying the detuning $\Delta_0$ at fixed coupling $\Omega_0$, as in Fig.~\ref{fig:delta-spectrum}.
The Efimov states will then show up as loss resonances whenever a bound trimer or tetramer state enters the Rabi-shifted continuum, similarly to how Efimov physics is typically probed using magnetic fields~\cite{Kraemer2006}. However, since the driving field is typically locked to an atomic frequency standard, this method of exploring Efimov states can potentially yield clearer signatures of highly excited states than in magnetic-field-based experiments. In order to address all the excited Efimov trimers up to the first in a CsLi mixture, i.e., to have $\Omega_0$ as large as $|E_0^{(1)}|$, we estimate that we need $\Omega_0 \lesssim 2\pi \times 100 \mathrm{kHz}$, which has recently been achieved in Ref.~\cite{vivanco2023}. We therefore expect that the modified three-body parameter predicted here can be observed in current or near-future experiments on strongly mass-imbalanced quantum gases.

While our work has focused on the behavior in the steady-state regime, there is also the prospect of using a strong Rabi drive to explore Efimov physics in the time domain, e.g., via Rabi oscillations. This could naturally be implemented for the Bose polaron scenario, where a small number of light impurities are immersed in a gas of heavy bosons. 
Thus, our few-body results also have potential ramifications for many-body systems such as Bose-Fermi mixtures.

\begin{acknowledgments} 
We gratefully acknowledge useful discussions with Olivier Bleu, Cesar Cabrera, Eleonora Lippi, Henning Moritz, Nir Navon, Michael Rautenberg, and Lincoln Turner. 
BCM, JL, and MMP acknowledge support from the Australian Research Council
Centre of Excellence in Future Low-Energy Electronics Technologies
(CE170100039).  JL and MMP are also supported through Australian Research Council Discovery Project DP240100569 and Future Fellowship FT200100619, respectively. ANZ acknowledges support from an Australian Government Research Training Program (RTP) Scholarship.
\end{acknowledgments}  

\paragraph{Data Availability.---} The data that support the findings of this article are openly available in the Figshare repository at~\cite{data}.

\bibliography{RabiEfimov}

\onecolumngrid
\newpage

\renewcommand{\theequation}{S\arabic{equation}}
\renewcommand{\thefigure}{S\arabic{figure}}
\renewcommand{\thesection}{S\arabic{section}}
\renewcommand{\thesubsection}{S\arabic{subsection}}
\renewcommand{\thetable}{S\arabic{table}}

\setcounter{equation}{0}
\setcounter{figure}{0}
\setcounter{table}{0}
\setcounter{page}{1}
\clearpage
\section*{SUPPLEMENTAL MATERIAL:\\ ``Universal Efimov Scaling in the Rabi-Coupled Few-Body Spectrum''}
\begin{center}
A. N. Zulli,
B. C. Mulkerin,
M. M. Parish,
and J. Levinsen\\
\emph{\small School of Physics and Astronomy, Monash University, Victoria 3800, Australia and}\\
\emph{\small ARC Centre of Excellence in Future Low-Energy Electronics Technologies, Monash University, Victoria 3800, Australia}
\end{center}

\section{Derivation of few-body equations}  
To derive the equations for the few-body bound states in the presence of Rabi-coupling, we start with the most general wave functions for the two-, three- and four-body systems, respectively given by:
\begin{subequations}
\begin{align}
    \ket{\Psi_{2}} &= \left(\sum_{\bk\sigma} \alpha^\pd_{\bk\sigma} \oppd{c}_{-\bk\sigma}\oppd{b}_{\bk} + \gamma^\pd_0\oppd{d}_{0} \right)\ket{0}\,, \label{eq:2wf}\\ 
    \ket{\Psi_{3}} &= \left(\frac{1}{2}\sum_{\bk_1\bk_2\sigma} \alpha^\pd_{\bk_1\bk_2\sigma} \oppd{c}_{ - \bk_1 - \bk_2 \sigma} \oppd{b}_{\bk_1} \oppd{b}_{\bk_2} + \sum_\bk \gamma^\pd_\bk \oppd{d}_{- \bk}\oppd{b}_{\bk} \right)\ket{0}\,, \label{eq:3wf}\\
    \ket{\Psi_{4}} &= \left( \frac{1}{6} \sum_{\bk_1\bk_2\bk_3\sigma} \alpha^\pd_{\bk_1\bk_2\bk_3\sigma} \oppd{c}_{ - \bk_1 - \bk_2 - \bk_3 \sigma} \oppd{b}_{\bk_1} \oppd{b}_{\bk_2} \oppd{b}_{\bk_3} + \frac{1}{2}\sum_{\bk_1\bk_2} \gamma^\pd_{\bk_1\bk_2} \oppd{d}_{ - \bk_1 - \bk_2} \oppd{b}_{\bk_1} \oppd{b}_{\bk_2}\right)\ket{0}\,.\label{eq:4wf}
\end{align}  
\end{subequations}
Here $\ket{0}$ is the vacuum state and $\{\alpha, \gamma\}$ form a set of amplitudes for each possible state. 
We then determine the bound-state energy $E$ by taking the \sch equation, $\hat{H} \ket{\Psi} = E \ket{\Psi}$, and projecting onto the different momentum states. 

For the two-body case in Eq.~\eqref{eq:2wf}, this gives the set of coupled equations
\begin{subequations}
\label{eq:2b-motion}
\begin{align}
    (E - \epsilon^b_\bk - \epsilon^c_\bk)\alpha_{\bk\up} &=  g\gamma_0 + \frac{\Omega_0}{2}\alpha_{\bk\down}\,, \\
    (E - \epsilon^b_\bk - \epsilon^c_\bk - \Delta_0)\alpha_{\bk\down} &= \frac{\Omega_0}{2}\alpha_{\bk\up}\,, \\
    (E - \nu_0)\gamma_0 &= g\sum_\bk \alpha_{\bk\up}\,.
\end{align} 
\end{subequations}
Solving for the amplitudes then yields an equation for the energy
\begin{equation}
    E-\nu_0 = g^2\sum_\bk\left(\frac{u^2}{E - \epsilon^b_\bk - \epsilon^c_\bk - \es{-}} + \frac{v^2}{E - \epsilon^b_\bk - \epsilon^c_\bk - \es{+}}\right)\, ,
\end{equation}
where $\es{\pm} = \big(\Delta_{0} \pm \sqrt{\Omega_0^2 + \Delta_{0}^2}\big)/2$,  and the real amplitudes $v$ and $u$ satisfy $u^{2} = \big(1 + \Delta_{0}/\sqrt{ \Omega_{0}^{2} + \Delta_{0}^{2}}\big)/2$ and $u^2 + v^2 = 1$. 
Finally, replacing the bare interaction parameters with the physical scattering parameters and performing the sum over momenta, we arrive at the following equation:
\begin{equation}
    \frac{m_{r}^2R^{*}}{\pi}E + \frac{m_{r}}{2\pi a} -\frac{m_{r}^{3/2}}{\sqrt{ 2 }\pi} \left(u^{2}\sqrt{\es{-} - E} + v^{2}\sqrt{\es{+} - E}\right) = 0\,.
\end{equation}
This exactly corresponds to the pole of the T matrix in Eq.~\eqref{eq:driven-two-body} of the main text, i.e., $\mathcal{T}_{\uparrow}^{-1}(E, 0) = 0$ in the center-of-mass frame, where $\k = 0$.

Similarly, Eq.~\eqref{eq:3wf} and Eq.~\eqref{eq:4wf} give, respectively, coupled equations for the 
three-body problem,
\begin{subequations}
\label{eq:3b-motion}
\begin{align}
    (E - \epsilon^b_{\bk_1} - \epsilon^b_{\bk_2} - \epsilon^c_{\bk_1 + \bk_2})\alpha_{\bk_1\bk_2\up} &=  g(\gamma_{\bk_1} + \gamma_{\bk_2}) + \frac{\Omega_0}{2}\alpha_{\bk_1\bk_2\down}\,, \\
    (E - \epsilon^b_{\bk_1} - \epsilon^b_{\bk_2} - \epsilon^c_{\bk_1 + \bk_2} - \Delta_0)\alpha_{\bk_1\bk_2\down} &= \frac{\Omega_0}{2}\alpha_{\bk_1\bk_2\up}\,,\\
    (E - \epsilon^d_{\bk} - \epsilon^b_{\bk} -  \nu_0)\gamma_\bk &= g\sum_{\bk'} \alpha_{\bk'\bk\up}\,,
\end{align} 
\end{subequations}
and the four-body problem,
\begin{subequations}
\label{eq:4b-motion}
\begin{align}
    (E - \epsilon^b_{\bk_1} - \epsilon^b_{\bk_2} - \epsilon^b_{\bk_3} - \epsilon^c_{\bk_1 + \bk_2 + \bk_3})\alpha_{\bk_1\bk_2\bk_3\up} &=  g(\gamma_{\bk_1\bk_2} + \gamma_{\bk_2\bk_3} + \gamma_{\bk_1\bk_3}) + \frac{\Omega_0}{2}\alpha_{\bk_1\bk_2\bk_3\down}\,, \\
    (E - \epsilon^b_{\bk_1} - \epsilon^b_{\bk_2} - \epsilon^b_{\bk_3} - \epsilon^c_{\bk_1 + \bk_2 + \bk_3} + \Delta_0)\alpha_{\bk_1\bk_2\bk_3\down} &=  \frac{\Omega_0}{2}\alpha_{\bk_1\bk_2\bk_3\up}\,, \\
    (E - \epsilon^d_{\bk_1 + \bk_2} - \epsilon^b_{\bk_1} - \epsilon^b_{\bk_2} -  \nu_0)\gamma_{\bk_1\bk_2} &= g\sum_{\bk'} \alpha_{\bk_1\bk_2\bk'\up}\,.
\end{align}  
\end{subequations}
Solving for $\alpha$ in terms of $\gamma$, and renormalizing the bare interaction parameters, finally yields the bound-state equations, Eqs.~\eqref{eq:three-body-T} and \eqref{eq:four-body-T}, in the main text.

\section{Rabi-driven Scattering Parameters}
To look at the effective scattering parameters in the presence of Rabi coupling, we transform from the spin basis to the Rabi-dressed basis and consider the T matrix for the lower Rabi-dressed state:
\begin{equation}
    \mathcal{T}_{\smallminus}(E) = u^2\mathcal{T}_{\uparrow}(E, 0)\,,
\end{equation}
where $\mathcal{T}_{\uparrow}(E, 0)$ is the spin-$\up$ T matrix defined in Eq.~\eqref{eq:driven-two-body} of the main text with zero center-of-mass momentum. 
For collisions with relative momentum $\k$, we can expand the standard low-energy scattering phase shift through 
\begin{equation}
    -k\cot{\delta} + ik = \frac{2\pi}{m_r}\mathcal{T}^{-1}_{\smallminus}(k^2/2m_r + \es{-}) \simeq \aef^{-1} + \rsef k^2 + ik\,,
\end{equation}
where we now have an effective scattering length, $\aef$, and a modified range parameter $\rsef$. This dressed scattering expansion is valid when we have $|\delta E| < \Omega_0$.
These scattering parameters are both functions of the Rabi-drive parameters and take the form:
\begin{equation}
    \begin{aligned}
        \aef &= u^{2}\left( \frac{1}{a} - \frac{1}{\ac} \right)^{-1} \,,\\
        \rsef &= \frac{1}{u^{2}}\left( \rs + v^{2} \frac{1}{2\sqrt{ 2m_{r}(\es{+} - \es{-}) }} \right)\,,
    \end{aligned}
\end{equation}
where $1/\ac = v^{2}\sqrt{ 2m_{r}(\es{+} -\es{-})} -2R^{*}\es{-}m_{r}$ is the (inverse) critical scattering length presented in Eq.~\eqref{eq:critical-scattering} in the main text. Here, we see that $\aef \to \infty$ when $1/a \to 1/\ac$, as expected, and that the range parameter is modified due to the coupling. In particular, we note that even when $R^*\to0$ we have $R^*_\text{eff} \neq 0$ in the presence of a Rabi coupling. However, as we discuss below, this Rabi-induced range parameter does not set a new three-body parameter, since this is a purely low-energy property, whereas the Efimov spectrum also relies on the behavior at short distances (high energies). 

\section{Model independence of few-body spectra}
In the main text, we used a two-channel model which naturally involves a finite effective range that impacts the two-body physics as well as providing an ultraviolet (UV) cutoff for the Efimov spectrum. To show that our key results are model independent, here we instead present all figures using an explicit UV cut-off that only affects the three- and four-body physics~\cite{yoshida2018, yoshida2018a}. To this end, we take the coupling $g$ to infinity while keeping the scattering length $a$ fixed in such a way that $R^* = 0$. In this limit, the two-body physics coincides with that of a single-channel model. To still provide a three-body parameter for the Efimov spectrum, we follow Ref.~\cite{yoshida2018} and introduce a UV cut-off $\Lambda$ on the momentum-exchange sums on the right-hand side of the few-body equations~\eqref{eq:three-body-T} and \eqref{eq:four-body-T} of the main text. 
The introduction of such a cut-off is equivalent to considering an explicit three-body repulsion \cite{bedaque1999}.
To distinguish the two models, we will denote the model with a three-body cut-off as the $\Lambda$ model, as opposed to the two-channel model considered in the main text, which we term the $R^*$ model. The deepest bound Efimov cluster for the \cs-\li mixture now enters the scattering continuum at $a_- = 179.578/\Lambda$.

\begin{figure}[hb!]
    \centering
    \includegraphics[width=\linewidth]{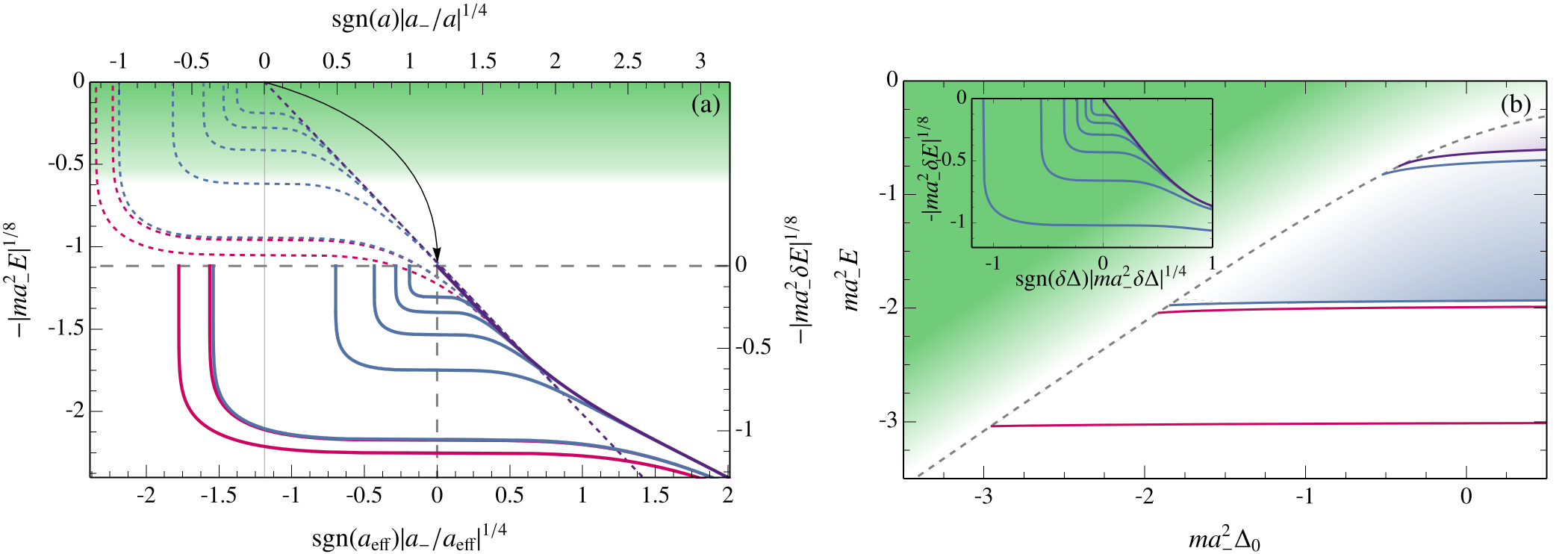}
    \caption{(a) The Efimov spectra  calculated within the $\Lambda$ model with (solid) and without (dashed) Rabi coupling. The two sets of axes represent the uncoupled case (left, top) and the Rabi-driven case (bottom, right), where the latter is expressed in units of the effective scattering length (see Eq.~\eqref{eq:scat-param} of the main text) and the shifted continuum ($\delta E = E - \es{-}$). The parameters are chosen to correspond to a \cs-\li mixture as in the main text, with $ma_-^2\Omega_0 = 2$ and $\Delta_{0}/\Omega_0 = -1$. The uncoupled Efimov spectrum (dashed) is centered around $(0, 0)$, while the coupled spectrum (solid) is shifted, and centered instead at $(a_\text{crit}, \es{-})$ (see arrow). The two-body bound state (purple), three-body trimers (blue) and four-body tetramers (pink) are visible in both spectra. (b) The few-body Rabi-coupled Efimov spectrum as a function of detuning, calculated within the $\Lambda$ model. The parameters correspond to a \cs-\li mixture as in the main text, with $ma_-^2\Omega_0 = 1$ and $a=|a_-|$. Here, we have shaded the scattering continuum (green), the continuum above the dimer (purple) and the continuum above the ground state trimer (blue). The two-body bound state (purple), trimers (blue) and tetramers (pink) all enter the continuum at the lower Rabi-coupled state, with energy $\es{-}$ (dashed line).  Inset: The Efimov spectrum centered around the critical detuning $\Dtwo$ as a function of $\delta\Delta = \Delta_{0} - \Dtwo$.}
    \label{fig:sm-lambda-spectra}
\end{figure}

In Fig.~\ref{fig:sm-lambda-spectra}, we plot the few-body spectra obtained within the $\Lambda$ model. Importantly, we adjust the parameters of the Rabi drive such that (when expressed in terms of $a_-$) they are the same as in the spectra in Figs.~\ref{fig:both-spectrum}(a) and \ref{fig:delta-spectrum} of the main text. 
A direct comparison of these figures reveals that the few-body physics within these two models is qualitatively identical, and thus that our key results are universal and model independent.

\begin{figure}[t!]
    \centering
    \includegraphics[width=0.5\linewidth]{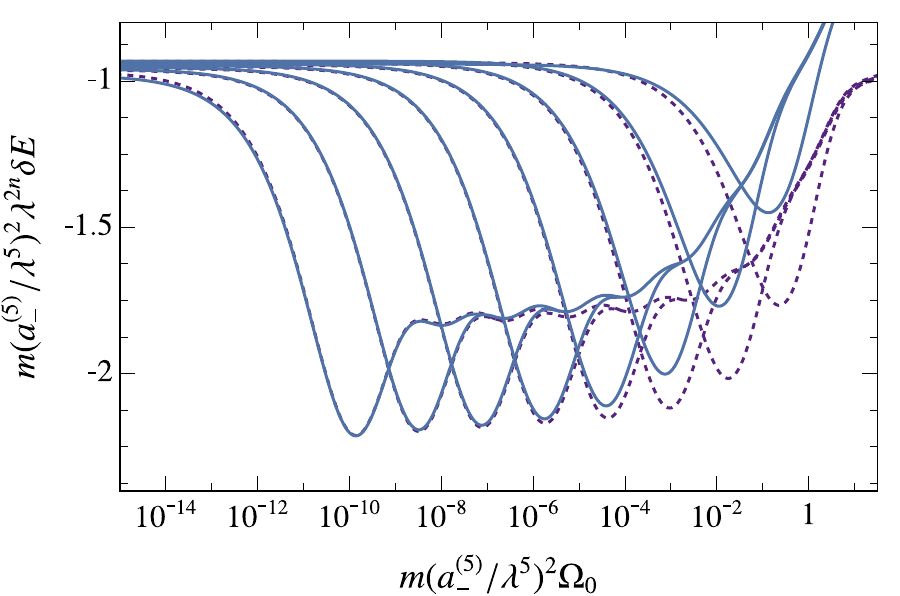}
    \caption{The energies of the first eight excited Efimov trimers at the induced resonance ($1/\aef=0)$, as a function of Rabi coupling. We show the comparison between the $\Lambda$ (purple, dashed) and $\rs$ (blue) models. We have defined $\delta E = E - \es{-}$ and multiplied it by the Efimovian scaling parameter $\lambda^{2n}$ to (nearly) collapse all states onto each other. To compare the spectra, we have parameterized them by $a_-^{(5)}$ which is far inside the universal regime.}
    \label{fig:E0-driveB}
\end{figure}

At a quantitative level, there are some differences between the results of the two models. This is primarily due to the fact that the effective range in the $R^*$ model modifies the two-body bound state when $a\lesssim R^*$, and thus (primarily) the deepest Efimov states. To check the universality in the regime where this dependence is negligible, we consider the Efimov states for $|a|\gg R^*$ in the $R^*$ model, where the two-body binding energy is essentially the same as in the $\Lambda$ model. The result is shown in Fig.~\ref{fig:E0-driveB}, similar to Fig.~\ref{fig:E0-drive}(a) in the main text, where we parameterize the Efimov states by $a_-^{(5)}/\lambda^5$ rather than by $\rs$. We see that the difference between the two models indeed disappears for the higher excited states. However, as we approach larger drives, and start probing higher energy physics, we again see the difference between the models, as expected. In both models, the new three-body parameter asymptotes to $0$ at very large drives, although the precise manner in which they asymptote is non-universal.

The deviations between the models can also be quantified by looking at the detunings at which the various Efimov states enter the scattering continuum for fixed Rabi drive and two-body scattering length (Fig.~\ref{fig:delta-spectrum} in the main text, and Fig.~\ref{fig:sm-lambda-spectra}(b) above). In the asymptotic limit of highly excited states, the ratio of consecutive critical values of $\Delta$ should approach the scaling parameter $\lambda=4.87$. In order to check the scaling parameter, we define the detuning values in relation to the new unitarity $\delta\Delta^{(n)} = \Dthr^{(n)} - \Dtwo$. Table \ref{tab:scaling} highlights the first few ratios for both models. As expected, the ratios of the first few states differ substantially between models, while the ratios for the highly excited states approach the universal value. 
 
\begin{table}[h!t]
    \centering
        \begin{tabularx}{0.5\linewidth}{Ycccccc}
        \hline
        \hline
            $\vphantom{\left(\displaystyle\frac{\delta\Delta^{(0)}}{\delta\Delta^{(1)}}\right)^{1}}$& $\displaystyle\frac{\delta\Delta^{(0)}}{\delta\Delta^{(1)}}$ & $\displaystyle\frac{\delta\Delta^{(1)}}{\delta\Delta^{(2)}}$ & $\displaystyle\frac{\delta\Delta^{(2)}}{\delta\Delta^{(3)}}$& $\displaystyle\frac{\delta\Delta^{(3)}}{\delta\Delta^{(4)}}$
           &$\displaystyle\frac{\delta\Delta^{(4)}}{\delta\Delta^{(5)}}$ & $\displaystyle\frac{\delta\Delta^{(5)}}{\delta\Delta^{(6)}}$   \\
        \hline
         $R^{*}$ model & $8.63$ & $5.57$ & $5.11$& $4.94$  &$4.89$ & $4.87$   \\
        $\Lambda$ model & $13.91$ & $6.40$ & $5.18$& $4.95$  &$4.89$ & $4.87$   \\
        \hline
        \hline
        \end{tabularx}
    \caption{Comparison of the detuning values at which the few-body bound states enter the scattering continuum. While the deepest bound states deviate from the universal asymptotic value of 4.87 \cite{pires2014, blume2014}, the highly excited states converge to this.}
    \label{tab:scaling}
\end{table}

\section{Born-Oppenheimer Approximation}
To gain insight into the essential physics that allows the three-body parameter to be tuned via the Rabi drive, we use the Born-Oppenheimer approximation~\cite{bethe1935}, which is valid in the experimentally relevant case of a \cs-\li mixture where $M \gg m$. In this limit, we can first assume that the light atom is moving around two fixed heavy atoms, located at $\RR_1$ and $\RR_2$. The $\sigma$ component of the Rabi-coupled light atom's wave function at position $\rr$ can then be written as $\psi_\sigma(\rr)$, which satisfies the non-interacting \sch equation everywhere 
\begin{align}
    \begin{pmatrix}  
        -\frac{\nabla_\rr^2}{2m_r} & \frac{\Omega_{0}}{2}\\  
        \frac{\Omega_{0}}{2} & -\frac{\nabla_\rr^2}{2m_r} +  \Delta_{0}  
    \end{pmatrix} \begin{pmatrix}
        \psi_{\uparrow} (\rr) \\
        \psi_{\downarrow} (\rr)
    \end{pmatrix} &= \epsilon(\RR_1, \RR_2)\begin{pmatrix}
        \psi_{\uparrow} (\rr) \\
        \psi_{\downarrow} (\rr)
    \end{pmatrix}\,,
\label{eq:non-interacting}
\end{align}
except at the positions of the heavy atoms where we impose the Bethe-Peierls boundary condition on the $\up$ component:
\begin{align}
     \psi_\uparrow(\rr) \propto \frac{1}{|\rr - \RR_i|} - \frac{1}{a} - 2m_r\rs \epsilon(\RR_1, \RR_2) &, \qquad \text{as} \quad |\rr - \RR_i| \to 0\,, \label{eq:bt-bc}
\end{align}
which includes the effective-range contribution~\cite{levinsen2011}. Here, %
we have used the reduced mass $m_r$ rather than $m$ in order to correctly describe the two-body physics. 
After solving Eqs.~\eqref{eq:non-interacting} and \eqref{eq:bt-bc}, we assume that the two heavy atoms move on the potential energy surface $\epsilon(\RR_1, \RR_2)$. Due to the translational symmetry of the system, we have $\epsilon(\RR_1 - \RR_2) \equiv \epsilon(R)$ with $R=|\RR_1 - \RR_2|$ the relative separation between the two heavy particles. To define an effective potential $\Vc(R)$ for the heavy particles, we must subtract the new scattering continuum, and thus $\Vc(R) = \epsilon(R) - \epsilon(\infty) = \epsilon(R) - \es{-}$, where $\es{-}$ is the shifted continuum. 
Then, the radial \sch equation for the relative motion of the two heavy particles satisfies the effective one-dimensional \sch equation 
\begin{equation}
\label{eq:SMsch}
    \left[- \frac{1}{M} \frac{\partial^2}{\partial R^2} %
    + \Vc(R)\right]\Psi(R) = \delta E\Psi(R)\,.
\end{equation}
Here, we have taken the $s$-wave case, since this is the relevant channel for Efimov bound states.

\begin{figure}[t]
    \centering
    \includegraphics[width=\linewidth]{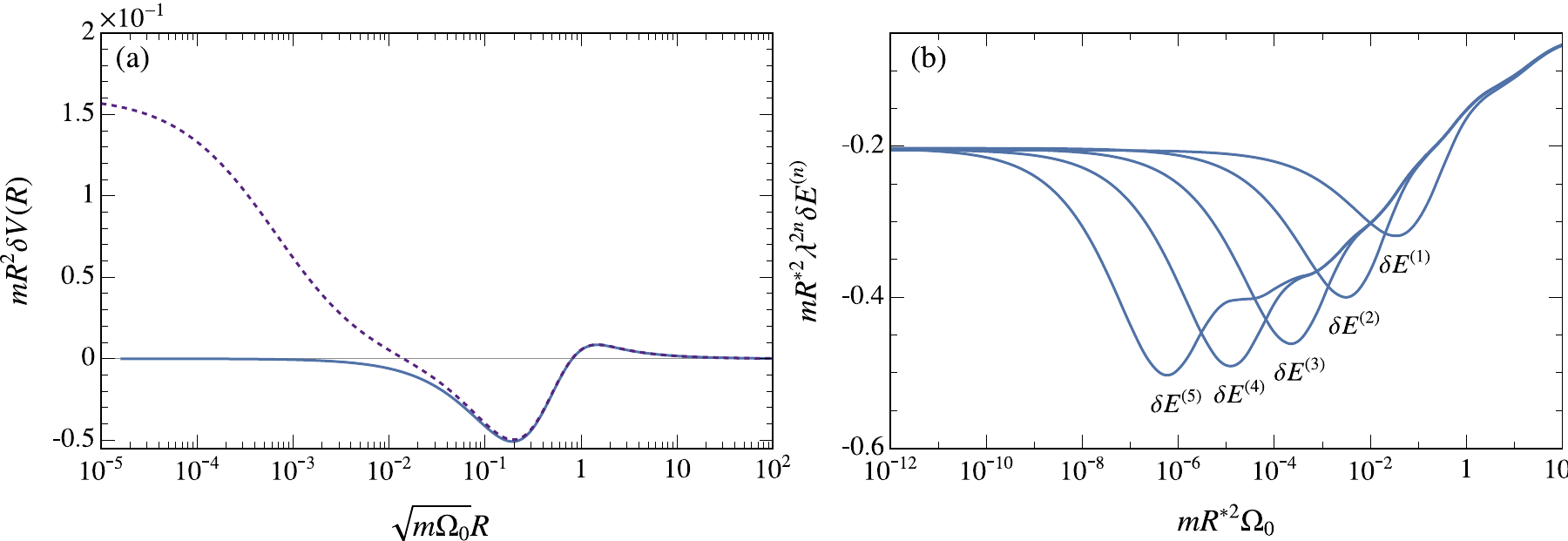}
    \caption{(a) Effective interaction potentials as a function of distance between the two heavy atoms, calculated within the Born-Oppenheimer approximation multiplied by $R^2$ to highlight the deviations from the original $-1/R^2$ potential. We plot the difference $\delta V(R)$ defined in Eq.~\eqref{eq:deltaV} at the induced resonance $1/\aef = 0$ and $\Delta_0 = 0$ with a finite effective range $\rs\sqrt{m\Omega_0} = 10^{-3}$ (purple, dashed) and zero effective range (blue).
    (b) Tuning the three-body parameter using Rabi coupling. We plot the energies of the first five excited Efimov states at the induced resonance with a finite effective range and $\Delta_0 = 0$ (purple dashed potential in panel (a)).}
    \label{fig:boFigs}
\end{figure}

To arrive at the effective potential, we need to solve Eqs.~\eqref{eq:non-interacting} and \eqref{eq:bt-bc}. As there are only interactions in the spin-$\uparrow$ state, the general solution can be written

\begin{equation}
    \psi_\uparrow(\rr) = f_1G^{(0)}_{\uparrow}(\rr, \RR_1) + f_2G^{(0)}_{\uparrow}(\rr, \RR_2)\,,
\end{equation}
which is similar to the case of the spin-orbit coupled few-body problem in Ref.~\cite{shi2015}. Here, $G^{(0)}_{\up}(\rr, \rr') = G^{(0)}_{\up}(|\rr - \rr'|)$ %
is the real-space Rabi-coupled Green's function for the spin-$\up$ light particle in the presence of one of the heavy particles,
\begin{equation}  
    G^{(0)}_{\up}(R) = -\frac{m_r}{2\pi}\frac{u^{2}e^{-\kapM R} + v^2e^{-\kapP R}}{R}\,,
\end{equation}
and $f_1$, $f_2$ are constants remaining to be determined. Since the system remains unchanged under exchange of $\RR_1$ and $\RR_2$ we have $f_1=\pm f_2$, with the minus sign resulting in the lowest energy $\epsilon$. 

We then apply the Bethe-Peierls boundary condition~\eqref{eq:bt-bc} and find the following matrix equation for $f_1$ and $f_2$:
\begin{equation}
    \begin{pmatrix}  
    \mathcal{T}_{\up}^{-1}\left(\epsilon(R), 0\right) & G^{(0)}_{\up}(R)\\  
    G^{(0)}_{\up}(R) & \mathcal{T}_{\up}^{-1}(\epsilon(R), 0)  
    \end{pmatrix} \begin{pmatrix}
    f_1 \\
    f_2
    \end{pmatrix} = 0\,,
\end{equation}
where we have the $T$ matrix, Eq.~\eqref{eq:driven-two-body} of the main text, %
\begin{equation} 
    \begin{aligned}
         \mathcal{T}_{\uparrow}^{-1}(E, 0) = \frac{m_r^2R^{*}}{\pi}E + \frac{m_r}{2\pi a} -\frac{m_r^{3/2}}{\sqrt{ 2 }\pi} \left(u^{2}\sqrt{\es{-} - E} + v^{2}\sqrt{\es{+} - E}\right)\,.
    \end{aligned}
\end{equation}
Finding the non-trivial solutions of this set of linear equations requires the determinant of the coefficient matrix to vanish. This leads to the following condition:
\begin{equation}
  \mathcal{T}_{\up}^{-1}(\epsilon(R), 0) = \pm G^{(0)}_{\up}(R)\,.
  \label{eq:SMdet}
\end{equation}
where we require the $+$ sign to have a physical solution.  

Of particular importance is the behavior at unitarity. In this case, in the absence of Rabi coupling and with $R^*=0$, the effective potential resulting from Eq.~\eqref{eq:SMdet} takes the particularly simple form~\cite{fonseca1979}
\begin{align} \label{eq:V0}
    V_0(R)\equiv -\frac{\W(1)^2}{2m_rR^2},
\end{align}
where $\W(x)$ is the Lambert $W$ function and $\W(1)\approx0.567$. This attractive $1/R^2$ potential is the origin of the Efimovian discrete scaling symmetry~\cite{efimov1970}, and we have $\lambda=e^{\pi/s_0}$ with $s_0=\sqrt{\W(1)^2M/2m_r-1/4} \simeq W(1)\sqrt{M/2m_r}$. 

The Rabi coupling modifies the effective potential, and in particular it shifts the position of the two-body resonance. In this case, Eq.~\eqref{eq:SMdet} at unitarity and $R^* = 0$ becomes
\begin{equation}
    \frac{e^{-\kappa R} + e^{-\sqrt{k_\Omega^2+\kappa^2}R}}R = \kappa + \sqrt{k_\Omega^2+\kappa^2} - k_\Omega \, ,
\end{equation}
where we have defined $\kappa = \sqrt{-2m_r\Vc(R)}$ and $k_\Omega = \sqrt{2m_r \Omega_0}$, and we have also taken a resonant Rabi drive such that $\Delta_0 = 0$, since we are interested in exploring universal behavior with a minimal number of energy scales. Thus, we now have an equation for the potential that depends on the Rabi coupling via the parameter $k_\Omega R$.

In the limits $k_\Omega R \to 0$ and $k_\Omega R \to \infty$, we see that the parameter $k_\Omega R$ vanishes from the equation and thus we expect to recover the potential in the absence of Rabi coupling, i.e., Eq.~\eqref{eq:V0}. This has two important consequences: First, it implies that the deepest bound states are essentially unmodified by the Rabi coupling, with a three-body parameter set by the usual UV physics. Second, we see that Efimov trimers also exist at arbitrarily small energies, with an unchanged Efimov scaling factor (but, possibly, a modified three-body parameter). The existence of the infinite tower of Efimov states in the Rabi-coupled system is a key difference compared with many of the previous scenarios considered, e.g., in the case where the light particle has two internal spin-orbit coupled states~\cite{shi2014, shi2015} it was found that the higher excited states were pushed into the continuum.

It is useful to quantify the difference between the Rabi-coupled potential at unitarity ($1/a_\mathrm{eff}=0$) and the universal form $V_0(R)$ as follows
\begin{align} \label{eq:deltaV}
    \delta V(R) =   %
    \Vc(R) + \frac{W(1)^2}{2m_rR^2}.
\end{align}
When multiplied by $R^2$, this resembles a scattering potential induced by the Rabi drive, as shown in  Fig.~\ref{fig:boFigs}(a). The appearance of a potential barrier at intermediate $R$ can be linked to the Rabi-induced range parameter $R^*_\text{eff}$, while the behavior at short distances is determined by the UV physics, such as the underlying range parameter $\rs$. Specifically, in the case of a finite $\rs$, the potential departs from the universal behavior in the limit $R \to 0$, corresponding instead to $\Vc(R)  \sim -\frac{1}{2m_r R\rs}$, yielding a repulsive barrier in $R^2 \delta V(R)$, as can be seen in Fig.~\ref{fig:boFigs}(a).

Indeed, following the approach of Ref.~\cite{sykes2017}, we can formally map this problem to a scattering problem by defining the new coordinate $z = \ln k_\Omega R$ and writing the wave function as $\Psi(R) = 
e^{z/2}\Phi(z)$. Then, in the regime where the energy is small, $|\delta E| \ll \Omega_0$, such that we can neglect it in Eq.~\eqref{eq:SMsch}, we arrive at the equation
\begin{equation}
    \label{eq:1d-scattering}
    - \frac{1}{M} \frac{\partial^2 \Phi}{\partial z^2} + \frac{u(z)}{2m_r} \Phi(z) = \frac{s_0^2}{M} \Phi(z) \, , 
\end{equation}
where $u(z) = W(1)^2-e^{2z}\kappa^2/k_\Omega^2$. 
Thus, we have a 1D problem of a particle with positive energy $s_0^2/M$ experiencing an effective potential $u(z)/2m_r$ that vanishes in the limits $z \to \pm \infty$ (corresponding to $k_\Omega R \to 0$ and $k_\Omega R \to \infty$).

The solution to Eq.~\eqref{eq:1d-scattering} can be written as 
\begin{equation}
    \Phi(z) = \sin(s_0 z + \delta(z)) \, ,
\end{equation}
where the phase $\delta(z)$ becomes constant away from the potential.
In the limit $z \to -\infty$, the phase is set by the short-distance boundary condition at some small $R_0$, i.e., $\delta(z) \to \delta_{-} = -s_0 \ln k_\Omega R_0$ and it is thus related to the (undriven) three-body parameter up to a constant scaling factor. 
Therefore, to determine the effect of the Rabi drive on the three-body parameter and the resulting Efimov spectrum, we simply need to calculate the phase shift due to the potential. In particular, for a given Efimov trimer, the ratio of trimer energies with and without the Rabi drive is given by $e^{2(\delta_+ - \delta_-)/s_0}$, where $\delta_+$ is the phase in the limit $z \to +\infty$.

According to the variable phase approach~\cite{calogero1967}, the phase satisfies the differential equation
\begin{equation} \label{eq:phaseshift}
    \frac{d \delta}{d z} = - \frac{M}{2m_r s_0} u(z) \sin^2(s_0z + \delta(z)) .
\end{equation}
The form of this equation immediately shows that the trimer energies have a log periodic dependence on the Rabi coupling $\Omega_0$, since it simply shifts the initial phase $\delta_-$ by a constant. 

In Fig.~\ref{fig:boFigs}(b) we show the Rabi-coupled spectrum calculated by numerically solving the \sch Eq.~\eqref{eq:SMsch} using the effective potential obtained from Eq.~\eqref{eq:SMdet} with the same parameters as in Fig.~\ref{fig:E0-drive}(a) of the main text. We see that the behavior of the effective three-body parameter closely resembles that calculated using the exact three-body equation, with minor differences arising from the use of the Born-Oppenheimer approximation.

\subsection{Three-body losses}
When three atoms approach each other---within distances of the order of $r_\text{vdW}$---they can recombine into a deeply bound dimer state. In this process, the molecular binding energy is released in the form of dimer-atom recoil kinetic energy, which is much larger than the trap depth and temperature. This process of three-body recombination is the main form of atom loss within resonant Bose gases and Bose-Fermi mixtures. %
We can model this loss process via an inelasticity parameter $\eta$~\cite{braaten2006, ulmanis2016a, sykes2017}, which corresponds to the addition of a small imaginary component to the phase $\delta$ and thus dictates the difference in amplitude between the incoming and outgoing waves. 

To investigate the effect of Rabi coupling on the inelasticity parameter, we once again employ the variable phase approach. Specifically, we consider a complex phase given by $\delta(z) = \delta_0(z) + i\eta(z)$, and solve Eq.~\eqref{eq:phaseshift}.  When $\eta(z) \ll \delta_0(z)$, Eq.~\eqref{eq:phaseshift} can be written  
\begin{subequations}
\begin{align} \label{eq:phaseshift-eta}
     \frac{d\delta_0}{dz} &\simeq -\frac{M}{2m_rs_0}u(z)\left[\sin^2\left(s_0z + \delta_0(z)\right) - \eta(z)^2\cos\left(2s_0z + 2\delta_0(z)\right)\right]\,,\\
    \frac{d\eta}{dz} &\simeq -\frac{M}{2m_rs_0}u(z)\eta(z)\sin\left(2s_0z + 2\delta_0(z)\right).
\end{align}
\end{subequations}
In the limit $z \to -\infty$, we have the modified short-distance boundary condition involving the underlying inelasticity parameter: $\delta(z) \to \delta_- + i\eta$.  Then, the phase to the right of the potential becomes $\delta(z) \to \delta_+ + i\eta_\Omega$, where $\eta_\Omega$ gives a measure of the Rabi-dressed loss rate, while the ratio of trimer energies is once again $e^{2(\delta_+ - \delta_-)/s_0}$.
From Eq.~\eqref{eq:phaseshift-eta}, we can see that the presence of the inelasticity parameter does not affect the log-periodic dependence on the Rabi drive, but it can change the amplitude of the oscillations in the trimer energies.

\begin{figure}[t]
    \centering
    \includegraphics[width=0.5\linewidth]{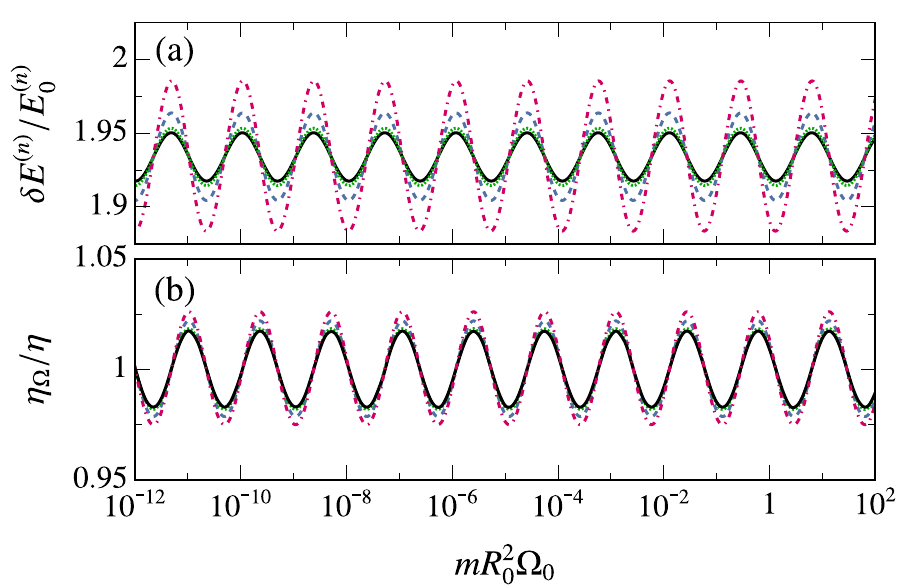}
    \caption{%
    Effect of three-body losses on the Rabi-dressed three-body parameter for the bare inelasticity $\eta \to 0^+$ (black, solid), $\eta = 0.3$ (green, dotted), $\eta = 0.6$ (blue, dashed) and $\eta = 0.9$ (pink, dot-dashed). For typical experimental parameters~\cite{ulmanis2016b}, this corresponds to $\eta \approx 0.6$ or $\eta \approx 0.9$. (a) The effect on the ratio of a given trimer's binding energy as a function of the Rabi coupling. (b) The Rabi-dressed inelasticity parameter as a ratio of the bare inelasticity parameter.}
    \label{fig:inelasticity}
\end{figure}

In Fig.~\ref{fig:inelasticity}(a), we show the effect of the inelasticity parameter on the ratio of the binding energy of a given trimer in the universal regime ($|\delta E| \ll \Omega_0$), with and without Rabi coupling, 
for various values of $\eta$. 
We find that while the average value of the ratio does not change, the amplitude of the log-periodic oscillations increases with larger inelasticity. Likewise, Fig.~\ref{fig:inelasticity}(b) shows the dressed inelasticity parameter, $\eta_\Omega$, which is similarly log-periodic with increasing Rabi coupling, with a mean value of the bare parameter. Therefore, we expect that the Rabi drive would have a minimal effect on the underlying collisional losses.

\subsection{Intraspecies Interactions}
For heteronuclear systems, the three-body parameter is not only dependent on the details of the pairwise interactions and the mass ratio, but can also depend on the intraspecies interactions \cite{haefner2017}. 
It is therefore interesting to investigate the manner in which such interactions can modify our results.

To this end, as in Ref.~\cite{ulmanis2016a, haefner2017}, we model the intraspecies interactions via a hard-core van der Waals potential, 
\begin{equation} \label{eq:hc-vdw}
    V_\mathrm{BB}(R) = \begin{cases} 
      \infty & R\leq R_0 \\
      -C_6/R^6 & R>R_0 
   \end{cases}\,,
\end{equation}
where $C_6$ is the van der Waals coefficient. Without the inclusion of the attractive van der Waals potential, the Cs-Cs scattering length is simply $a_\mathrm{BB}=R_0$.
In general, $a_\mathrm{BB}$ is related to  $R_0$ and $C_6$ via the equation~\cite{gribakin1993,tran2021}
\begin{equation} \label{eq:acscs}
    \frac{Y_{1/4}(2R_\mathrm{vdW}/R^2_0)}{J_{1/4}(2R_\mathrm{vdW}/R^2_0)} = 1 - \sqrt{2}\frac{a_\mathrm{BB}}{r_\mathrm{vdW}}\frac{\Gamma(5/4)}{\Gamma(3/4)},
\end{equation}
where $J_\alpha(x)$ [$Y_\alpha(x)$] are the Bessel functions of the first [second] kind, $\Gamma(x)$ is the Gamma function and $R_\mathrm{vdW} = 1/2 (MC_6)^{1/4}$ is the van der Waals length. Equation~\eqref{eq:acscs} has multiple solutions for a given $a_\mathrm{BB}$, corresponding to different numbers of bound states in the attractive potential. Therefore, the specific value of $R_0$ not only sets $a_\mathrm{BB}$, but also the total number of supported bound states. Following the work of Refs.~\cite{ulmanis2016a,haefner2017}, we choose $R_0$ such that the potential supports %
two Cs dimer states. 

In the absence of Rabi coupling, it is known that for small positive $a_\text{BB}$ and negative heavy-light interactions, the deepest bound Efimov trimers do not enter the three-body continuum \cite{haefner2017, ulmanis2016a}. Instead, they dissociate into the %
light atom plus heavy-heavy dimer continuum, causing the three-body parameter to be drastically shifted from the models without Cs-Cs interaction. Therefore, we now include the intraspecies interactions in the Rabi-coupled CsLi three-body system using the effective potential $\Vc$ and the intraspecies interactions in Eq.~\eqref{eq:acscs}.

\begin{figure}[t]
    \centering
    \includegraphics[width=0.5\linewidth]{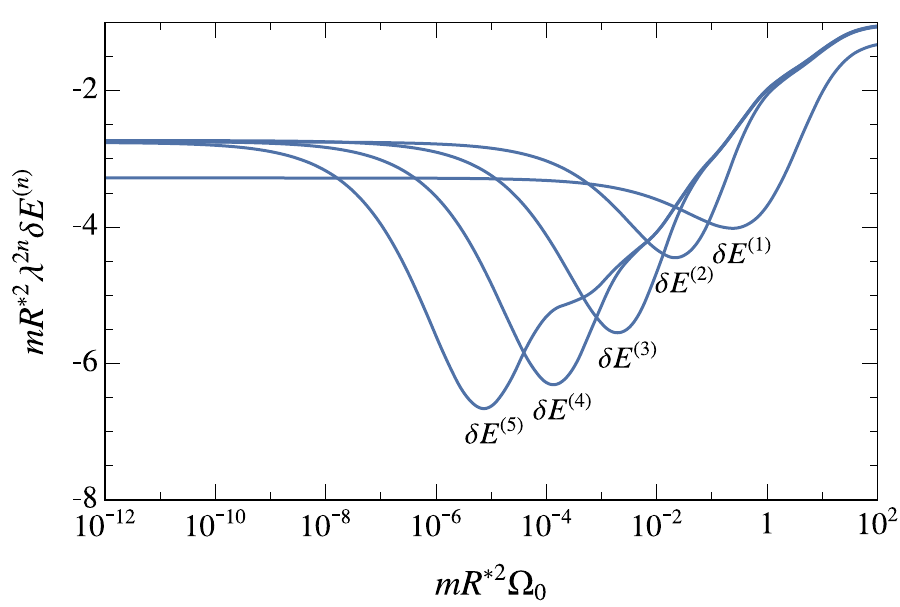}
    \caption{Energies of the first five excited Efimov states at the induced resonance for $\Delta_0 = 0$. Here, we have included intraspecies (Cs-Cs) scattering, with a scattering length $a_\mathrm{BB} = R^*/2$. For typical experimental parameters~\cite{tung2014}, this corresponds to $a_\mathrm{BB} \approx 100a_0$.}
    \label{fig:cscsinteractions}
\end{figure}

Within the Born-Oppenheimer approximation, our effective BB \sch equation is modified to be: 
\begin{equation}
\label{eq:SMschCsCs}
    \left[- \frac{1}{M} \frac{\partial^2}{\partial R^2} %
    + \Vc(R) +  V_\text{BB}(R)\right]\Psi(R) = \delta E\Psi(R)\,.
\end{equation}
Solving Eq.~\eqref{eq:SMschCsCs} allows us to capture the interplay between the three relevant scales, namely $\Omega_0$, $R^*$ and $a_\text{BB}$. In Fig.~\ref{fig:cscsinteractions} we perform the same analysis as in Fig.~2(a) of the main text and Fig.~\ref{fig:boFigs}(b) above, where now we look at the behavior of the Rabi-dressed energies as a function of increasing drive strength for the specific value $a_\text{BB} = \rs/2$. %
As expected, the ground state is now so deeply bound that it %
is completely insensitive to the drive. %
On the other hand, evidently the excited Efimov states look qualitatively very similar to those in Fig.~\ref{fig:boFigs}(b), albeit with a different energy in the limit $\Omega_0\to0$, indicating that (as expected) the system has a modified three-body parameter. Crucially, the relative energy shifts due to Rabi coupling are essentially exactly the same as in the case where $a_\mathrm{BB}=0$, indicating that the Rabi drive modifies the three-body parameter in precisely the same manner. Therefore, we see that no matter what sets the three-body parameter, the Rabi-dressing universally shifts the original three-body parameter. This justifies the omission of $a_\mathrm{BB}$ from the results presented in the main text.

\section{Tuning the three-body parameter}
The ability to modify the effective three-body parameter by a strong Rabi drive is a key result of our work. Here, we investigate this further by showing how a large range of possible three-body parameters can be explored by varying the detuning. We also consider the effect of changing the mass ratio of the two species.
\begin{figure}[t]
    \centering
    \includegraphics[width=\linewidth]{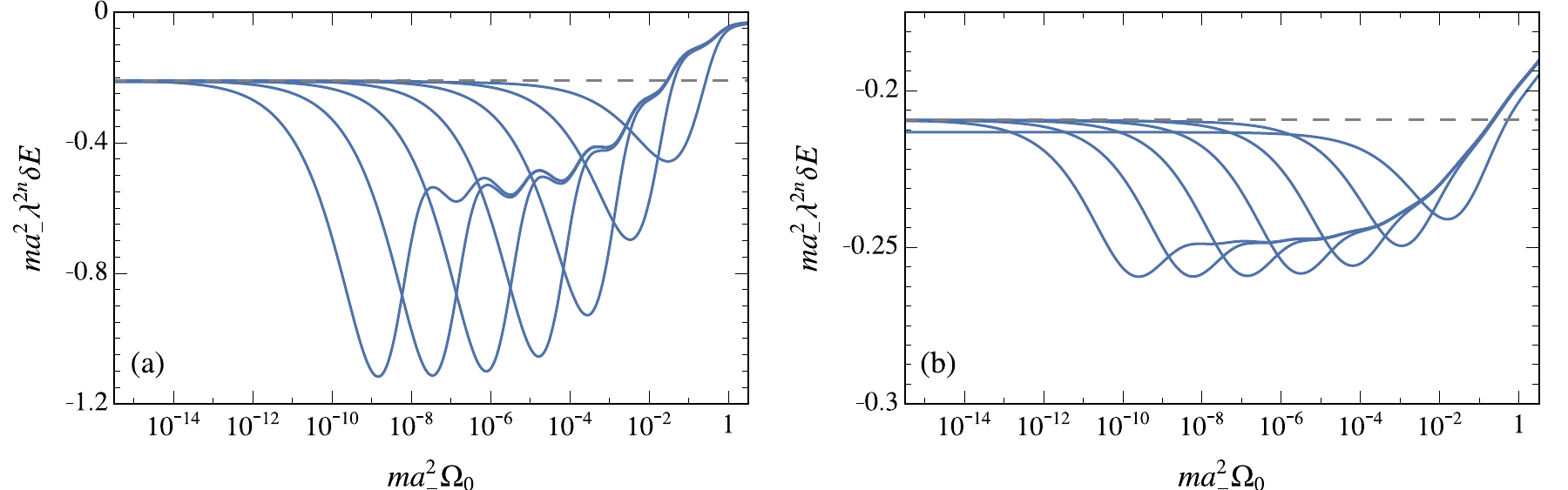}
    \caption{The first seven excited Efimov trimer energies at the induced resonance ($1/a_{\mathrm{eff}}=0$) as a function of Rabi-coupling strength, calculated with the $\rs$ model. Here, we have applied the scaling relation $E_0^{(n)}\lambda^{2n} = E_0$ to collapse all the binding energies to a single initial value. The light gray dashed lines represent the condition $\es{-} = E_0^{(n)}$, where the rightmost is the ground $n=0$ case. In (a) we have $\Delta_0=-\Omega_0$ and in (b) $\Delta_0=\Omega_0$.}
    \label{fig:E0Detunings}
\end{figure}
In Fig.~\ref{fig:E0-drive}(a) of the main text, we presented the three-body parameter as a function of Rabi coupling at zero detuning. In Fig.~\ref{fig:E0Detunings} we keep all parameters the same except the detuning which is chosen to be equal in magnitude to the coupling, either positive or negative (this ensures that the fractions $u^2$ and $v^2$ of the spin components remain constant as we vary $\Omega_0$). We find that in the case of negative detuning, the states become much deeper bound before they turn to follow the universal curve, which is now set by both $\Delta_0$ and $\Omega_0$. Notably, the energies on the universal curve have been shifted by almost a factor 3 compared with the case in the absence of Rabi coupling, which, given the scaling ratio of $\lambda=4.87$, means that we can explore a large fraction of the possible effective three-body parameters.  
By comparison, with positive detuning the states only show a small change in the effective three-body parameter. This effect mirrors the values of $1/u^2$ in each of these regimes, namely $1/u^2 \approx {6.8, 2, 1.1}$ for negative, zero, and positive detuning respectively.

\begin{figure}[b]
    \centering
    \includegraphics[width=0.5\linewidth]{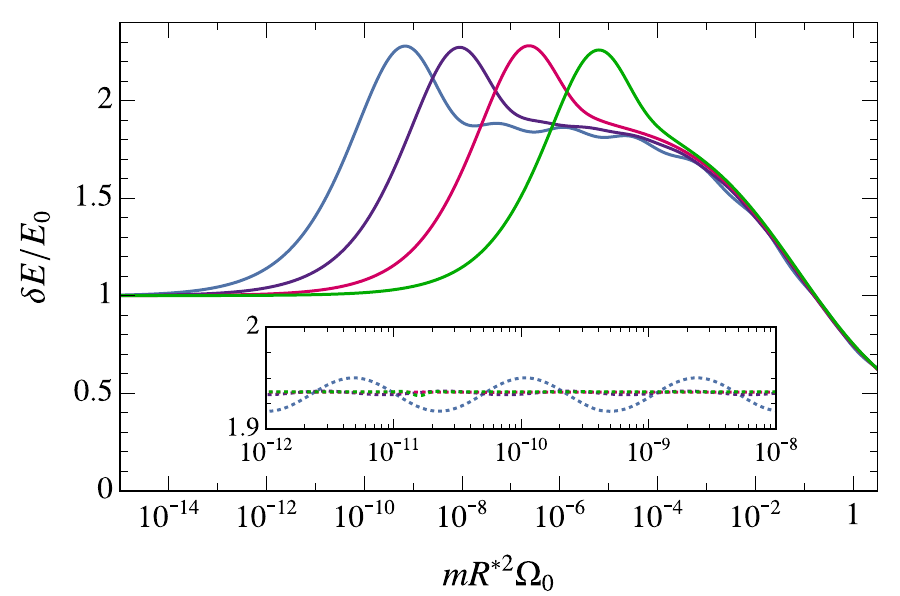}
    \caption{The coupled binding energies at the induced resonance as a ratio of the initial binding energy for various mass ratios, calculated with the $\rs$ model and $\Delta_0 = 0$.  Here, we show the 7th state for $M/m \approx 22.1$ (blue), 9th state for $M/m \approx 44.3$ (purple), 11th state for $M/m \approx 88.6$ (pink), 13th state for $M/m \approx 177.6$ (green). The scaling factor for these mass ratios are: $4.87, 3.05, 2.17$ and $1.72$ respectively. Inset: The ratio of coupled to uncoupled energies calculated within the Born-Oppenheimer approximation using the variable phase approach, Eq.~\eqref{eq:phaseshift}, for the same mass ratios.}
    \label{fig:massScaling}
\end{figure}

Finally, Fig.~\ref{fig:massScaling} shows the binding energy of a highly excited trimer for four different mass ratios.  
This illustrates how the energy ratio is nearly independent of mass ratio, even towards the non-universal regime of large Rabi drive. On the other hand, the oscillations as a function of Rabi drive $\Omega_0$ die down for large mass ratios (see inset).

\end{document}